\documentclass[10pt]{article}
\usepackage{graphicx}
\usepackage{longtable}
\usepackage{times}
\usepackage{setspace}
\usepackage{dcolumn}
\usepackage{bm}
\usepackage{rotating}
\usepackage{hyperref}
\usepackage{epstopdf}
\usepackage{bm}
\usepackage{rotating}
\begin{document}
%\begin{doublespace}
\begin{center}
    \textbf{The effect of porosity of dust particles on polarization and color with special reference to comets}
\end{center}

\begin{center}
    $ A. \ K. \ Sen $
\end{center}
\begin{center}
    \textit{Department of Physics, Assam University, Silchar-788001, India \\ $ asokesen@yahoo.com $}
\end{center}
\begin{center}
    $ R  \ Botet $
\end{center}
\begin{center}
    \textit{Laboratoire de Physique des Solides UMR8502, Universit\'e Paris-Sud, Universit\'e Paris-Saclay, Orsay 91405, France \\ $ robert.botet@u-psud.fr $}
\end{center}
\begin{center}
    $ R. \ Vilaplana $
\end{center}
\begin{center}
    \textit{Dpto. de F\'isica Aplicada, Universidad Polit\'ecnica de Valencia, E.P.S.A.  Alcoy 03801, Spain\\ $ rovilap@fis.upv.es $}
\end{center}
\begin{center}
    $ Naznin \ R \ Choudhury $
\end{center}
\begin{center}
    \textit{Department of Physics, Assam University, Silchar-788001, India \\ $ nazninrahimchoudhury@gmail.com $}
\end{center}
\begin{center}
    $ Ranjan \  Gupta $
\end{center}
\begin{center}
    \textit{IUCAA, Ganeshkhinde, Pune 411007, India \\ $ rag@iucaa.in $}
\end{center}
\begin{center}
Date: 3 October 2016
\end{center}
\begin{abstract}
 Cosmic dusts are mostly responsible for polarization of the light that we observe from astrophysical objects.
They also lead to color-extinction, thermal re-emission and other scattering related phenomena. Dusts are made of small particles which are characterised by their size (radius), composition (matter), and  structure (morphology, including porosity). In the present work, we address the question of the role of the dust particle porosity on light polarization and color, using Discrete Dipole Approximation (DDA) light scattering code.
To answer this question, we developed an algorithm to generate solid particles of arbitrary values of porosity. In brief, the model considers a given homogeneous structure made of touching dipoles. The dipoles are randomly removed one by one, such that the remaining structure remains connected. We stop the removal process when the desired porosity is obtained. Then we study the optical properties of the porous particle.
That way, we show how the proper value of the porosity affects the polarization and color of the light scattered by these porous particles. In addition to polarization, porosity has important effects on photometric color.
Considering an important application, we emphasize the possible role of the porosity of the cometary dust particles on polarization and color of the light scattered by cometary coma.
\end{abstract}
\section{Introduction}

Ever since the first polarization measurements have been made for astronomical objects, scientists are using  polarimetry as a tool to probe astrophysical processes in celestial environments. It has been the case for the comets since the first observation of the great comet C/1819 N1 tail polarization by Arago (von Humboldt 1858) and his conclusions that comet was not a self-luminous object and that its polarization was the result of reflection of the sunlight on material lost by the comet. It has also been the case for the interstellar dust, evidenced independently by Hall (1949) and Hiltner (1949) from their measurements of  stars polarization that was interpreted as starlight scattered by elongated small particles aligned in the magnetic field (Greenberg 1951).

Nowadays, among the various processes that may cause polarization from remote astronomical sources, light scattering by small solid particles is expected to be the most common case. Such cosmic dust may appear naturally under the form either of dispersed particles (comet tails, reflection nebula, etc.) or of deposited  particles (regoliths, pond-like deposits, etc.). Information about shape and composition of the cosmic dust particles should come from theory, which tells us how the light polarization is related to the various  particle properties and to the scattering geometries. \\

%In this paper, we shall use the word \emph{dust} or \emph{aggregate} to mean one and the same thing,  which is the astrophysical dust particle responsible for light scattering.

Historically, the theory of light polarization from dusts should start from the successful explanation of the sky polarization by Rayleigh (1871). Indeed, at a time where the concept of molecules was still unclear (Maxwell 1873), Rayleigh studied explicitly light scattered by minute solid particles in order to skirt questions about the nature of the atmospheric particles.
Thus, the Rayleigh dipolar scattering theory applies fully to any particle much smaller than the wavelength and of refractive index close to 1 -- possibly the case for dusty media in the infrared domain. Later, Mie  proposed in 1905 the analytical solution of the scattering of a plane electromagnetic wave by a perfect homogeneous dielectric sphere of any radius. This solution (van de Hulst 1957) provided a real breakthrough in the field, since it allowed for the first time to tackle the relation between polarization from dusts and  astrophysical particle sizes (in all the following, ``particle size'' means the typical radius of the particle, thus we consider particles which are not essentially anisotropic).

Generally, dust particles are not expected to be perfect spheres, then the Mie theory is only approximation of the real process -- though a very important one indeed.   The advent of the computer in the 1970s  has allowed to deal with light scattered by other shapes and structures.
Over the years, numerical tools were made available to compute light scattered by irregularly shaped  particles which were  expected to be most naturally occurring in astrophysical conditions. Such numerical simulation codes include T-matrix (Mishchenko et al. 1994), DDSCAT (Draine 2000), FDTD (Taflove \& Hagness 2005), to mention just a few.

Over the years, as we have understood  that astrophysical dusts should be made of disordered porous / fluffy particles, scientists came up with various numerical codes to simulate such dust particles. That way, algorithms were developed to simulate formation processes of  irregularly shaped  dust particles under realistic physical conditions. The most popular code, the Cluster-Cluster Aggregation mode (Jullien \& Botet, 1987) and its variants, provide reliable numerical analogues for astrophysical dusts made of aggregates of small grains. Such aggregates are often fractal -- hence of fluffy structures, though they may exhibit irregular compact shapes as well, depending on the conditions.

There is a set of commonly accepted terminologies that is worth recalling here, since it may depend on the domain of physics where dust particles are considered. In the context of astrophysics, a ``grain'' is tiny solid particle much smaller than a characteristic length scale (in the present context: much smaller than the wavelength). A grain is generally assumed to be spherical in shape because the surface irregularities of such a minute solid are irrelevant.
An absence of a grain in a compact particle will introduce porosity. We use the words \emph{fluffy} or \emph{porous} with the same intention to mean voids in the aggregates.

Finally, a grain can be represented by a collection of dipoles to comply with the Discrete-Dipole Approximation (DDA), which is the base of the DDSCAT calculations (Draine 2000). The dipoles are the electromagnetic counterparts of nanoscopic compact domains of  standard shape, occupying a definite position in a three dimensional regular lattice structure and having complex refractive index assigned to each individual dipole (Shen et al. 2008, 2009). \\

Porous / fluffy aggregates are expected to exist everywhere in outer space, from our solar system to the interstellar medium and also in other galaxies. However, in this paper the results that we obtain from our simulations will be compared with the data from comets. Indeed, there is huge amount of data for comets observed at various scattering geometries, and the ground-based data are complemented by in situ space measurements. This provides a very stable and large data base to compare with the simulated results.

Existence of such porous / fluffy particles in comets, in interplanetary medium and in our upper atmosphere was confirmed by various in situ (Brownlee 1985) and space measurements. Their existence was also predicted by  astronomers from the  analysis of ground-based data (Greenberg \& Hage 1990; Xing \& Hanner  1997;  Kimura  2001;  Kimura  et  al.  2006  etc.). Both  laboratory  and  computer  simulations  of  solar  radiation  scattered  from cometary  dusts  have  been  carried  out to  understand  the  structures  of cometary  dust particles  (Wurm  \&  Blum  1998;  Gustafson  \&  Kolokolova 1999; Hadamcik  et  al.  2010,2013,2015;  Xing  \&  Hanner  1997;  Kimura 2001; Petrova, Tishkovets \& Jockers 2004; Tishkovets, Petrova \& Jockers 2004; Gupta et al 2006; Kimura, Kolokolova \& Mann 2006; Bertini, Thomas \& Barberi 2007; Kolokolova et al. 2007; Moreno et al. 2007;  Das \& Sen 2011; Das et al. 2011 and others). \\

In most of the astrophysical conditions, the gases are so diluted that grain aggregation is expected to follow the Ballistic Aggregation procedure (Meakin 1983, 1984), that is all the particles and grains move along random straight trajectories. According to the physcial conditions, the process can be either successive sticking of grains on aggregates (this is the Ballistic Particle-Cluster Aggregation model, BPCA), or diffusion is similar for all the particles and the aggregates grow essentially sticking together (this is the Ballistic Cluster-Cluster Aggregation model, BCCA). The BPCA aggregates are compact in the fractal sense (that is fractal dimension of the BPCA aggregates is equal to 3), while the BCCA aggregates are fractal with the fractal dimension $d_f \simeq 1.9$. Laboratory diagnostics of particle coagulation in the solar nebula suggests that the particles grow preferentially under the BCCA process (Wurm \& Blum 1998). A variant of the BCCA is the Reaction-Limited Cluster-Cluster Aggregation model (RCCA). In the RCCA, sticking is hard to occur, for example because the aggregates are electrically charged. Then, two fractal particles may randomly collide and interpenetrate because of their fluffy structures, but without sticking because of the electric charges distributed on their grains. When by chance entanglement is strong, the aggregates may stay long together, and after some time, the relative kinetic energy of the entangled aggregates allow to overcome the repulsive forces and it results in effective contact, that is sticking by van der Waals forces. This process leads to aggregates of fractal dimension $d_f \simeq 2.1$.  In astrophysical clouds, BCCA is expected to hold when the grain electric charges are irrelevant, and RCCA when the grain electric charges are relevant. \\

As we are here more interested in compact shapes, we may recall that Shen et al.(2008,2009) considered interesting modifications to standard BPCAs in which local grains migrate along the surface of the aggregates by rolling or sliding over the first-contact grain. These reorganizations produce even more compact aggregates than BPCA. The models are called Ballistic Aggregation with Migration (BAM).  Allowing for one  (BAM1) or  two  (BAM2) migrations  produce aggregates of increasing density. The BAM models are not based on realistic rearrangement of the grains inside the particles, and they are used as ``cheap approach'' (Seizinger \& Kley 2013) to generate aggregates denser than those obtained using the realistic BPCA process.

Other aggregation models were proposed for specific purposes. A famous example is the off-lattice Eden model, used to simulate tumor growth. In this model, grains are added randomly on any location tangent to a grain of the aggregate and realizing no overlaps with other grains. It results in very compact structure with irregular surface.

Unlike the fractal particles, the compact-shape particles are not characterized by their fractal dimension (since it realizes the maximum value, $d_f = 3$), but by their porosity, that is the fraction of voids inside the particle. For instance, large BPCA aggregates porosity (fraction of void inside the particle) is about $85 \%$, whereas the BAM2 aggregates are denser with porosity about $70 \%$.  The porosity of Eden-model particles is about $35 \%$ (Wang et al 1995).

\section{On the definition of the porosity}

As a starting point of the problem, we consider the list, by order of importance, of the parameters relevant for characterizing the particle scattering profile.

The most important  parameter is the total volume of the particle, that is the amount of matter interacting with the electromagnetic wave. This parameter was already highlighted in the Rayleigh theory, and has be written in various forms, worth mentioning under the form of the effective radius,  $a_{eff}$,
defined by the equality between the total volume of the matter inside the particle, $V_m$, and the volume of the perfect sphere of radius $a_{eff}$, that is:
\begin{equation}
V_m = \frac{4}{3} \pi  a_{eff}^2 \label{vm}
\end{equation}

Porosity is another parameter ubiquitous in the physics of the disordered small particles, but it is not yet clear whether the parameter is relevant or not for the light scattering process. The present study tries to put ideas in order about that quantity.
The first question is the proper porosity definition. Actually, one should consider two cases:

\begin{itemize}
\item
the closed porosity is the case where the particle can be defined as a compact volume with voids inside it. In this context,  voids are called: `pores'. The porosity, $f$, of a particle of volume $V_m$ is naturally defined as:
\begin{equation}
f = \frac{V_p}{V_m+V_p} \label{f}
\end{equation}
with $V_p$ the total volume of the pores. Note that $f$ depends only on the non-dimensional ratio $V_p/V_m$ between the total volume of the pores and the volume of the matter inside a given geometric volume.

Condensation of droplets in which gas bubbles are dispersed gives an example of such porous particles. This is the case for ash produced by coal pyrolysis (Jiang 2006).

\item
the open porosity is the case where the outer surface irregularities are of  same order as the overall size of the particle. Such open structures going from surface to inner places are sometimes called `fjords' by reference to the long, narrow, inlets going into the coasts of Norway (Barab\'asi \& Vicsek 1990). A generic example is the case of fractal particles. If $d_f < 3$ denotes the fractal dimension of the matter distribution inside the particle, then the total volume of the matter inside a sphere of radius $R$ smaller than the overall radius of the particle, scales as $ \simeq c (R/a_o)^{d_f}$, with $c$ a numerical constant and $a_o$ the typical size of the grains the particle is made of. Then, generalizing the concept of pores to the general voids inside a meaningful geometrical sphere, the porosity, as defined by Eqn. (\ref{f}), writes:
\begin{equation}
f = 1 - c \left( \frac{R}{a_o} \right)^{d_f -3} \label{ff}
\end{equation}
Taking $R$ as the typical radius of the particle, one notes that the porosity Eqn. (\ref{ff}) is size-dependent, and goes to $1$ for the large particles since $d_f-3 <0$. This is in opposition to the fractal dimension, $d_f$, which is characteristic of the process and not of the size of the particles.
\end{itemize}

Despite the embarrassing possible size-dependence of the porosity, one can argue that $f$  is a relevant parameter for  light scattering. Indeed, $V_m$ is the amount of homogeneous matter surrounded by empty space, interacting with the light, and $V_p$ is similarly the amount of voids surrounded by homogeneous matter,  interacting with the light (from Babinet's principle). Then, using $V_m$ and $f$ appears to be the most important shape parameters to categorize light scattering of finite particles.

The next step (not considered in the present work) would be to consider the various correlations either in the matter or in the void distribution.

\section{Models and Methods}

In the following, we will discuss the case of the closed porosity, leaving for a companion article the case of the open porosity.

\subsection{Generation of random particles with closed porosity}

As our knowledge about cometary dust becomes more complete thanks to extensive observations and in-situ cometary missions, the very high-porosity of the particles ejected by comet nucleus, appears to be essential result. A simple way to model very porous particles is to use the Ballistic Cluster-Cluster Aggregation model (Wurm \& Blum 1998). This approach may be relevant as the BCCA model corresponds to  general physical conditions possibly met in the cometary environment (Meakin 1983, 1984). However, we have to be aware of a possible flaw: forcing the fractal particles to realize a given value of the particle porosity requires to fix the average radius of the particles, because of Eqn. (\ref{ff}).

Here, we take the opposite point of view: we do not mind the fractal structure of the particles, but we fix the porosity of the particles to a given value. Taking a model of closed porosity, the value of $f$ is a constant whatever the size of the particle is. In other words, $V_m$ and $V_p$ are the two independent parameters of the model.

This class of model corresponds to the following process:  outer layer of the comet nucleus becomes more and more porous through  mechanisms which may include release of trapped gases (including gas sublimation), irradiation by solar radiation, interaction with solar wind, etc. Therefore, microscopic voids form naturally down to the penetration depth of  heat and solar radiations.  The dust particles are formed in a second process from the breakup of the porous  layer by the gas escaping through the inner parts of the nucleus.

\subsection{Generation of the porous dust particles}

Considering the DDA method to obtain  light scattering properties, one starts from a large number of electric dipoles regularly placed on the nodes of  a cubic lattice. Two dipoles are said to be connected if they are nearest neighbors. The first ensemble of $N$ dipoles is arranged initially in the shape of a sphere of radius $a$. Voids are generated according to the following procedure: one dipole is chosen randomly and removed. If this change leaves the ensemble of the remaining dipoles all connected (that is every couple of dipoles can be linked by a sequence of connected dipoles), one keeps the change. If it is not the case (removing that dipole leads to isolated dipole or isolated group of dipoles), then  removal is canceled, the dipole is put back in its original position. Then, one removes another dipole. The process stops when the desired porosity is obtained. Previously, similar studies were carried out with the dipoles unconnected and hence, such aggregates were unrealistic (Vilaplana et al. 2006 and 2011).

The test for connection of the ensemble of dipoles is realized through epidemic algorithm (Demers et al 1987). More precisely: a dipole is chosen randomly. Then, the list of all the dipoles connected to it is calculated (infection). One considers all the dipoles in the list and do the same calculation, that is determining all the dipoles, not considered before, and connected to them (spreading).  At the end of the process, one knows the list of all the dipoles connected to the very first dipole. If all the dipoles of the actual ensemble are in the final list, the ensemble is  connected. Otherwise the ensemble is disconnected. \\

This model allows to generate solid spherical particles of definite porosity, taking into account cohesion of the structure. These particles are homogeneous in average because of the random removal process. At the end, most of the pores are totaly included inside the structure. Other pores impinge on the particle surface, but they are a few in number and not relevant (unlike the `fjords' of a fractal structure). According to the definition by Eqn.  (\ref{f}), the final ensemble of dipoles contains $(1-f)N$ dipoles. A slice of a particle of porosity $f = 50\%$ is shown in the Fig.1.

\begin{figure}
	% To include a figure from a file named example.*
	% Allowable file formats are eps or ps if compiling using latex
	% or pdf, png, jpg if compiling using pdflatex
	%\includegraphics[width=\columnwidth]{pc50-1.eps}
    \includegraphics[width=\columnwidth]{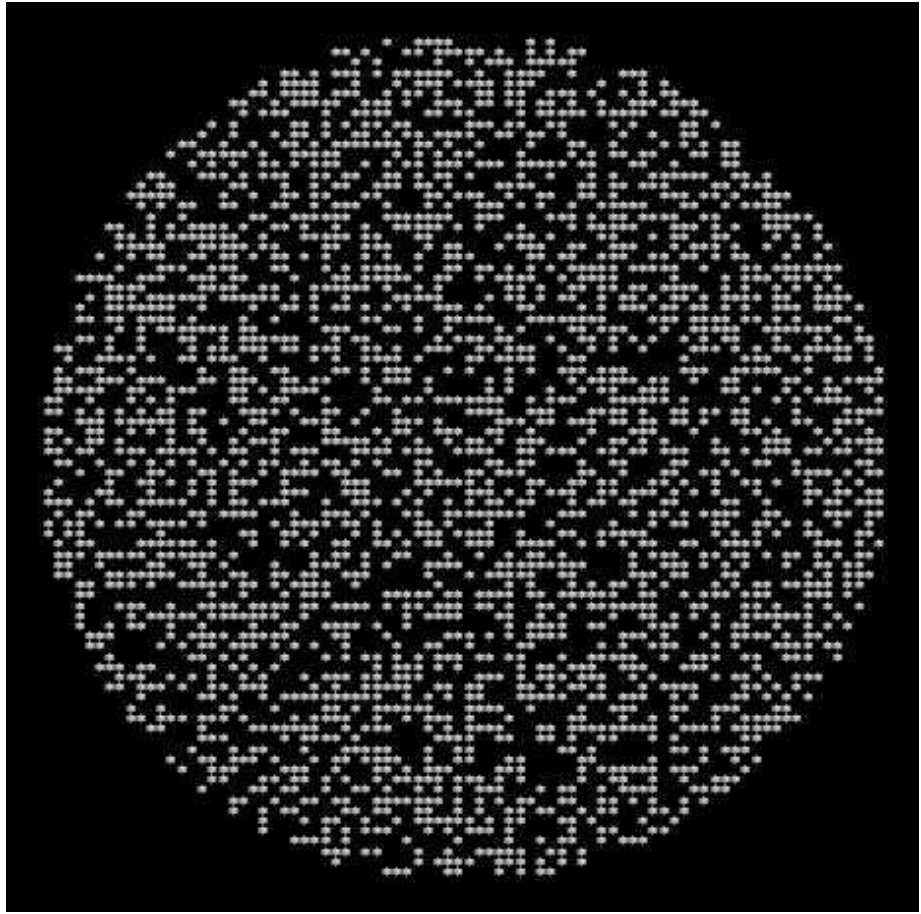}
    \caption{Slice of an aggregate structure generated by our code for the 50\% porosity value. The width of the slice is 1 grain-diameter. The slice passes through the geometric center of the particle.}
    \label{por50_3.8}
\end{figure}

\subsection{Use of DDA code and generation of light scattering different matrix elements for randomly porous aggregate}

In order to calculate the  polarization and intensities of scattered radiation, we use DDSCAT 7.3 light scattering code. We find that this code has the desired level of flexibility in order to be applied on the porous aggregate structures
that we have reported in the section above.
Even if our approach is quite general and can be used similarly in many different situations, we will concentrate on applications to the cometary dust particles since large and reliable data base is currently available.

\subsubsection{range of the particle diameters}

The dust particles ejected from comets have a wide range of sizes. We take here the lower limit for the particles diameter: $0.01 \mu$m -- that is the lower limit used for  Halley and other types of comets in the past (Sen et al. 1991).  Ideally the upper limit for size distribution can go up to several micrometers, according to conclusions from ground-based observations of recently studied comets, such as 67P (Hadamcik et al 2011, 2013, 2015).
Also, for the comet 67P, the on-board GIADA instrument during ROSETTA mission had detected the presence
of large particles  (Rotundi et al. 2015), as well as many porous dust particles.

For computational reasons, we limit here the upper bound of particle size distribution to $1.0 \mu$m, in accordance with many other authors (Das et al. 2011 and references therein). Indeed, the size distribution is often represented as an inverse power law, $\nu(a) \sim 1/a^{\alpha}$ with large  index $\alpha$ values ranging from $1.8$ to $3.8$ (Das et al 2011 ; Kolokolova et al 1997,2001 ; Mukai et al 1987 ; Sen et al 1991 ). Then, the very large particles are so infrequent that they are expected not to influence the averaged light scattering properties.

In our analysis below, we shall use  the following values of the size-distribution exponent, $\alpha$, namely: $\alpha = 1.8$, $2.8$ and $3.8$, in order to  check how the simulated values can match with the observed polarization data for comets.
\subsubsection{composition of the particles}

For the composition of cometary dusts, we choose pyroxene with complex refractive index values as $(1.692,0.0492)$ at the wavelength $0.485 \mu$m, and $(1.672, 0.0185)$ at the wavelength $0.684 \mu$m (Das et al. 2011), with the justification for using this material given in Section 4. A large data base on cometary polarization for P/Halley is available at these two wavelengths.

\subsubsection{computed optical parameters}

Running the DDSCAT 7.3  code allows to generate the Mueller matrix elements  $S_{11}, S_{12},$ $S_{13}$ etc. needed to compute all the optical quantities, including  light polarization (Bohren and Hoffman 1983 ; Draine and Flatau 1994).
The calculations are done performing averages over $10 \times 10 \times 10$ random orientations, and  by integration over the dust size distribution.
The linear polarization, $P$, is calculated using the relation:
\begin{equation}
P = -\frac{S_{12}}{S_{11}} \label{p0}
\end{equation}

In principle, one should take into account the $S_{13}$ coefficient, using instead (Bohren and Hofman 1983):
\begin{equation}
P = \frac{\sqrt{S_{12}^2+S_{13}^2}}{S_{11}} \label{p1}
\end{equation}

However,  the random aggregates generated by our code have spherical symmetry in average, then we expect to have: $\langle S_{13} \rangle = 0$ when averaged over angular randomization and the definition by Eqn. (\ref{p0}) can be used as a good approximation.

More precisely, difference between Eqns. (\ref{p1}) and (\ref{p0}) is a factor $ \sqrt{1+(S_{13}/S_{12})^2}$.
In our present simulations, the highest $S_{13}/S_{12}$ value, that we encountered is $9\%$. This has  occurred   for the $50\%$ porosity particle at $\lambda=0.485 \mu$m and with power law index  $\alpha=3.8$.
This $S_{13}/S_{12}$ value should influence the estimated polarization value by a multiplicative factor of $\sqrt{1+(0.0899)^2}=1.004$. We are very much aware from observational astronomers that the uncertainty in observed polarization is never below $0.5 \%$, and thus a factor of $0.4 \%$ will have no effect on our simulation and modeling work. We also checked on some examples that both (\ref{p0}) and (\ref{p1}) gave very similar values of the polarization.
However, under a different situation, the linear polarization should be expressed by (\ref{p1}).

\section{Results and Discussions}

In recent years Vilaplana et al ( 2006, 2011) had addressed this problem of effect of porosity  on the polarization ($P$) and light scattering efficiency ($Q_{sca}$).
The authors generated particles with fluffiness/porosity by uniformly randomly removing dipoles, without constraining particles to remain connected.
They defined the fluffiness degree as the percentage of dipoles
removed and generated aggregates with  fluffiness degrees as $15 \%, 25 \%$ and $50 \%$. As can be seen in the Figs. 9, 10 and 11 of Vilaplana et al.(2011), a displacement to the right and rising of the first maximum of $Q_{sca}$ is observed as the fluffiness degree increases. The
result is that both the real and imaginary parts of the \emph{effective} refractive index decrease simultaneously as the fluffiness
degree is increased. This behaviour can be expected as the authors calculated the effective refractive index of the porous dust by Maxwell-Garnet (1907) theory.

In the present work, we consider connected aggregate structures with various porosities. We start with compact structures where we have 12,0000 dipoles and then we generate  structures
with various porosity values $10 \%, 20 \%, 30 \%, 40 \%$ and $50 \%$. While running DDSCAT code, we do averaging over $10 \times 10 \times 10$ random orientations. We use the power law size distribution
with lower bound and upper bound as $0.01$ and $1.0 \mu$m as discussed earlier. We perform integration over this size range numerically, by doing the calculation in 71 steps and then using Simpson's 1/3 rule. The index for power law dust size distribution is considered as 1.8, 2.8 and 3.8 as these are the most commonly used
values in literature (Das et al 2011 and references therein). As far the compositions  are concerned, there can be many possible options, and scientists often
consider mixtures of various types. But since we want here to focus our studies on $porosity$, we restrict ourselves to only one type of composition which is pyroxene
marked by the complex refractive index values of (1.692, 0.0492) and (1.672, 0.0185)   at the wavelengths $\lambda = 0.485 \mu$m and 0.684 $\mu$m respectively (Das et al. 2011).
This material is most commonly used as a candidate for cometary dust.

\subsection{Effect of porosity  of aggregates on polarization }

In this section we discuss  the effect of porosity, $f$, on  cometary polarizations, with the porosity defined generally through Eqn. (\ref{f}).

It was noticed  (Das et al 2011) that the linear-polarization curve of the Halley's comet is bell-shaped -- which  is  a general feature among all the comets --,  with max-value, $P_{max}$, and min-value, $P_{min}$, about $25 \%$ and $-5 \%$ respectively.
However, these values are not very accurate, since it was not possible to make polarization measurements at all the scattering angles.
 We should also note that $P_{max}$  can be even higher than $25\%$ and $P_{min}$ even more negative than $-5\%$.
Crossover from positive to negative polarization values takes place at a scattering angle $\simeq 158$ degrees.

In the Figs. 2- 7, we draw simulated polarization  curves at various porosities and various power law indices of the particle size distribution. We find that with the increase in porosity (from $f = 0\%$ to $f = 50\%$) both the maximum and minimum  polarization values change. The  negative polarization branch becomes deeper with the increase in porosity and then again it becomes shallower. This  non-monotonic phenomenon appears for both the wavelengths $0.485 \mu$m and $0.684 \mu$m, and with the three values of power law index $\alpha = 1.8, 2.8$ and 3.8.

Among all the different sets of parameters considered here, the matching with the observed polarization data of comet is better obtained with 40\% porosity and  $\alpha = 3.8$ power law index at both the wavelengths.
In Table I, we list only those simulated polarization values which are close to the above observed polarization values of $P_{max}$, $P_{min}$ and consistent with crossover angles.

We could even fine tune the porosity and power law index values, and try to match the observed polarization data more closely.
By a $\chi ^2$ minimization technique, one can  find the optimum values of porosity ($f$) and power law index ($\alpha$), which realize the best fit to the observed polarization data. This we postpone for future, which will be more detailed work. \\

Way back in 1987, immediately after the apparition comet Halley, Mukai et al (1987), carried out simulation work by varying the refractive index (that is composition) of cometary dust. As one can see from the Fig. 1 of their article, various shapes for the polarization versus scattering-angle curve are possible,
as we change the refractive index of cometary dust. In the present work, we can see in a similar way that various shapes for  the polarization versus scattering-angle curve are possible by changing the porosity $f$.
These two results may be related since an effective refractive index can be found dependent on the actual porosity using the Maxwell-Garnett effective medium theory. However, this procedure may be somewhat empirical and it will be discussed in more details below in the Section 5. On the other hand, the procedure developed in the present work, provides a definite and analytical way of dealing with porous dust.

\subsection{Effect of the porosity on the color of the aggregates}

In addition to polarization features, comets also show interesting properties of photometric \emph{color}. Since, there are many cometary molecules having their own emission bands within the visible region, we have to observe the comets only through some specific wavelength bands, where we can measure the light scattered by cometary dust alone and uncontaminated by cometary molecular emissions.

The ratio of the scattered intensities as observed through such two continuum bands, when expressed in suitable log scale is called \emph{color}. Kolokolova et al (1997) had worked on cometary color and elaborated on the definition of the concept of cometary color.
As stated by the authors, the \emph{color}, $ C$, can be suitably  defined as  :
\begin{equation}
 C= -2.5 \log \frac {I(0.485)}{I(0.684)} ~.  \label{D}
\end{equation}
The quantity $I(0.485)$ and $I(0.684)$ are the two recorded intensities of light scattered by cometary dusts through the two wavebands centered around $0.485 \mu$m and $0.684 \mu$m respectively.
In other words,  cometary \emph{color} can be measured simply by the relation:
\begin{equation}
C= B(0.485)- R(0.684) ~,
\end{equation}
where the B(0.485) and R(0.684) are the cometary magnitudes observed through the above cometary filters, which were
standardized by International Halley Watch (IHW) and widely used for cometary photometry.

The  \emph{color}, $C$, can be estimated theoretically using Eqn. (\ref{D}) by considering certain specific cometary dust population. The detailed procedure has been described by Kolokolova  et al (1987). From their Equations (9) and (10), we arrive at the following two equations below, which will be applicable directly in our case:

\begin{equation}
 I(0.485)= \int^{amax}_{amin}{S_{11}(0.485,a) \frac{d a}{a^{\alpha}}} ~. \label{I1}
\end{equation}
%\begin{equation}
% I(0.485)= \int^{amax}_{amin}{S_{11}(0.485,a)\frac{da}{a^{\alpha}}} ~.  \label{I1}
%\end{equation}

\begin{equation}
I(0.684)= \int^{amax}_{amin}{S_{11}(0.684,a) \frac{d a}{a^{\alpha}}} ~. \label{I2}
\end{equation}
where $S_{11}$  is the total intensity scattering matrix element obtained after running DDSCAT code at the given wavelengths $ 0.485 \mu$m and $0.684 \mu$m and for the given dust radius $a$ expressed in $\mu$m. The dust size follows the power law distribution  $\nu(a) \sim 1/a^{\alpha}$.
 The lower and upper limit of dust size distributions are $a_{min}$ and $a_{max}$  are respectively $0.01 \mu$m and $1.0 \mu$m in the present case. \\

Here, the  \emph{color} values are calculated from the simulation data using the Eqs. (\ref{D}), (\ref{I1}) and (\ref{I2})  for different porosity values of cometary dust. The \emph{color}s are plotted against the scattering angles on the Figs. 8, 9, 10 for different power law indices.
Looking at these figures, we find that the \emph{color} values, $C$,  remain almost constant for different scattering angles at a given porosity, except sudden increase in the domain of the negative branch of the polarization curve (that is for scattering angles > 160 degrees). Still more remarkable, the constant value of the \emph{color} with the scattering angle does not change appreciably for the different porosities at a given power index value. One can conclude from these remarks that the color of the cometary dust depends essentially on the size-distribution of the dust particles and not on their porosity, with the noticeable exceptions for the backward  scattering directions.
This is confirmation of previous conclusions by Kolokolova et al (1997). Indeed, Kolokolova et al. (1997, 2001) using power law index $\alpha=  3$ or still higher, concluded that the cometary $color$ is mainly determined by the size distribution of the cometary dust, whereas the cometary polarization is mainly determined by the complex refractive index. \\

However, we cannot explain quantitatively at the same time both the polarization curve of the comet Halley  and its \emph{color}. Indeed, the average $color$ of the comet is known to be  $C \simeq -0.024$ (Kolokolova et al 1997), which is a value much smaller than the $color$ found in the simulations for the best fit of the polarization curve (namely: $C \simeq -1$ for porosity $40\%$ and power law index $\alpha = 3.8$).

Actually, as the estimated $color$ values are the $\simeq -0.2$, $\simeq -0.6$ and $\simeq -1.0$ respectively for the three power law index values $\alpha = 1.8, 2.8$ and 3.8, then the $color (C)$ is always found to be much more negative than the observed Halley's comet $color$. A possible interpretation of the numerical data is that big dust particles (more numerous for the small values of $\alpha$) tend to be white, whereas small-size particles (dominant for the large values of $\alpha$)  are essentially reddish.
Therefore, the number of large particles may well have been underestimated compared to the real Halley comet dust size-distribution. Larger value of the size cutoff $a_{max}$ should be tempted, but that means adding a free parameter to the model, then making the simulations and their analysis much more lengthy. By the way, let us  note that, till today, there have been no successful attempts where a single dust model could  explain both  polarization and $color$.

\section{Relevance of the Effective Medium Theory }

As mentioned in the previous Sections and also discussed by Maxwell-Garnett (1907), the effect of porosity, $f$, of dust particles can be approximately replaced by replacing the actual matter refractive index $n=(m,k)$ for an equivalent homogeneous matter of refractive index $n'= (m',k')$. Such approach is  known as \emph{Effective Medium Theory}. Various schemes are available for determining effective refractive index.  Following Voshchinnikov et al (2007), we adopt here the mixing rule of Bruggeman, since it has been found to work for wide porosity range and deviations to the exact results do not exceed  $5\%$.

This symmetric mixing rule is stated below with our set of symbols :
\begin{equation}
(1-f)\frac{n^2- n'^2}{n^2+2 n'^2}+ f \frac{1-n'^2}{1+2n'^2}=0 ~.
\end{equation}
Solving this complex-value equation allows to compute the effective refractive index for any set of the parameters $\{f, n, k \}$. Then, using the Mie result for the homogeneous spheres with radii distributed as $\nu(a)$, approximate averaged polarization and \emph{color} of the ensemble of porous particle characterized by  $\{f, m, k \}$ and $\nu(a)$ can be calculated. This scheme was used previously for polarization (\emph{viz}. Sen et al. 1991). \\

The values of the effective refractive index are shown in the Table 2.
The corresponding polarization and \emph{color} values are plotted along with those previously obtained using porous aggregates in the Figs. 3 - 11. From these figures it is clear that  the Effective Medium Theory can not exactly reproduce  the numerical polarization and color values obtained using DDA method on the simulated porous aggregates, though most of the data are comparable. The differences become more noticeable, as  the porosity increases. Vaidya et al (2007) had also observed similar limitations of Effective Medium Theory.

\begin{figure}
	% To include a figure from a file named example.*
	% Allowable file formats are eps or ps if compiling using latex
	% or pdf, png, jpg if compiling using pdflatex
	\includegraphics[width=\columnwidth]{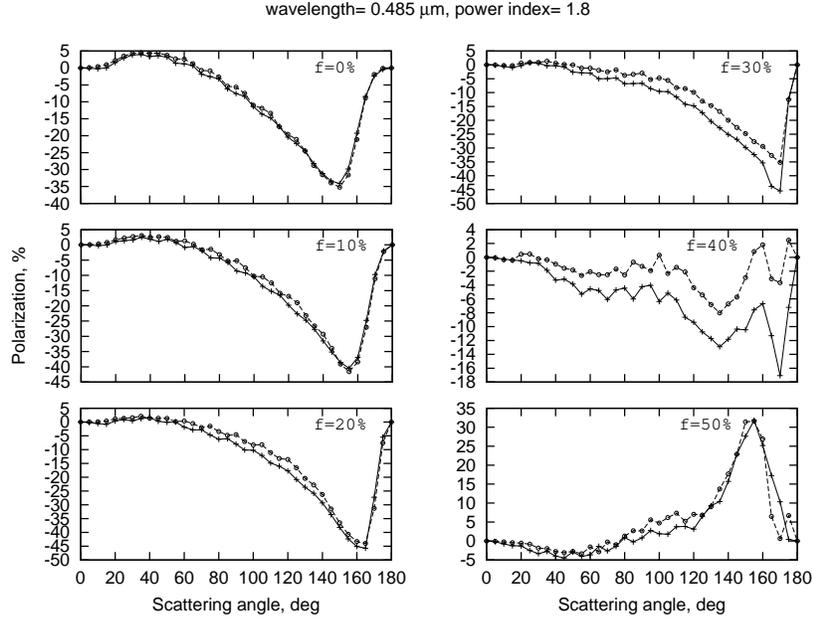}
    \caption{The above figures show the variations of dust polarization with scattering angles expressed in degrees for various values of dust particle porosities.
    The polarization has been calculated for particles having size range $ 0.01 \mu$m - $1.0 \mu$m, with a power-law size  index 1.8 and  wavelength $ \lambda$  =0.485 $\mu$m. The  curve marked by 'o' is the one drawn  with the DDA method, and the  curve  joined by '+' is the one drawn with the Mie theory.}
    \label{pol-485-1.8}
\end{figure}

\begin{figure}
	% To include a figure from a file named example.*
	% Allowable file formats are eps or ps if compiling using latex
	% or pdf, png, jpg if compiling using pdflatex
	\includegraphics[width=\columnwidth]{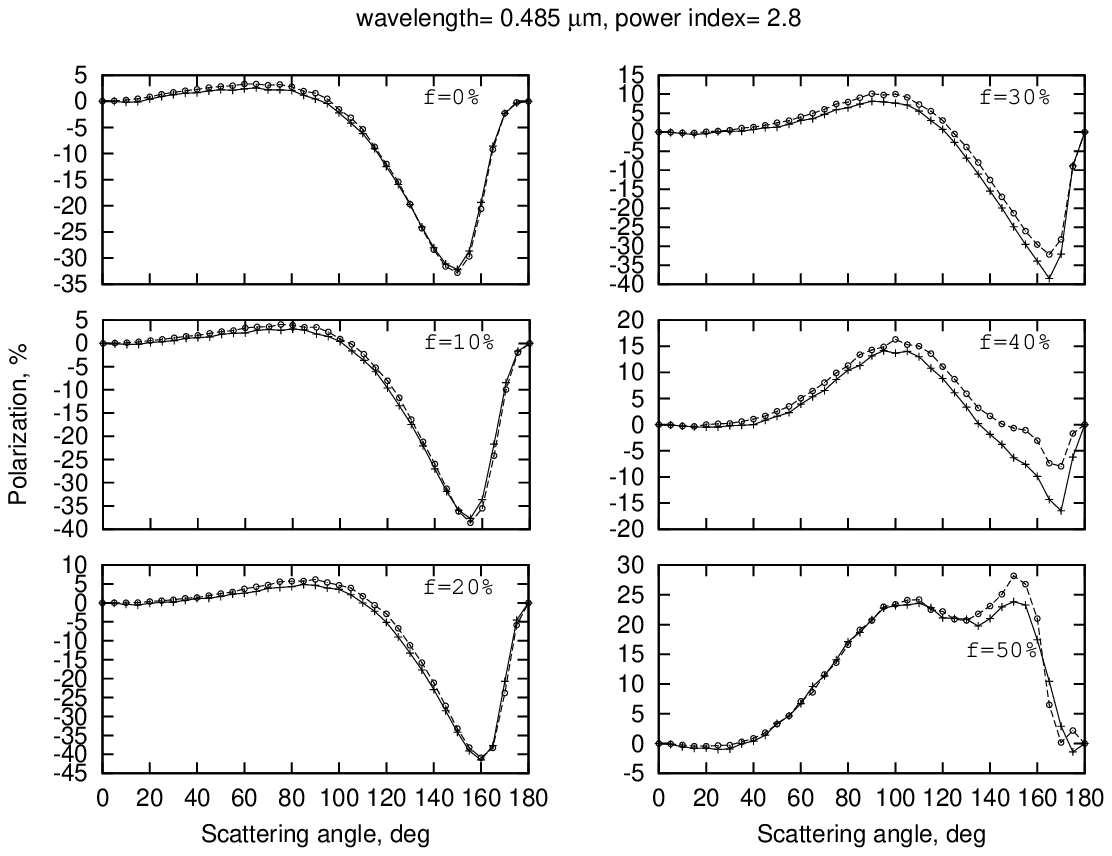}
    \caption{The above figures show the variations of dust polarization with scattering angles expressed in degrees, for various values of dust particle porosities.
    The polarization has been calculated for particles having size range $ 0.01 \mu$m - $1.0 \mu$m, with a power-law size  index $\alpha = 2.8$ and  wavelength $\lambda$  = 0.485 $\mu$m. The  curve (marked by 'o' symbols) is the one drawn  using the DDA method, and the  curve  (joined by '+' symbols) is the one drawn using the Mie theory.}
    \label{pol-485-2.8}
\end{figure}

\begin{figure}
	% To include a figure from a file named example.*
	% Allowable file formats are eps or ps if compiling using latex
	% or pdf, png, jpg if compiling using pdflatex
	\includegraphics[width=\columnwidth]{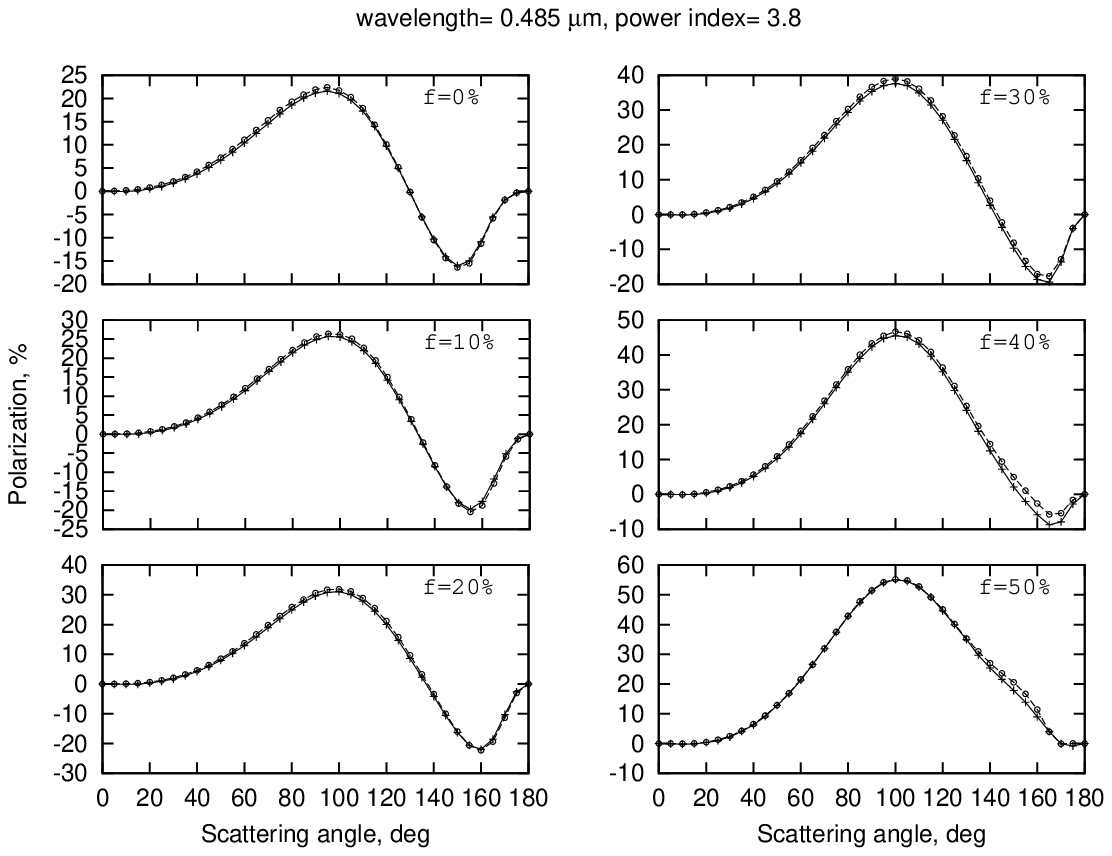}
    \caption{The above figures  the variations of dust polarization with scattering angles expressed in degrees, for various values of dust particle porosities.
    The polarization has been calculated for particles having size range $ 0.01 \mu$m - $1.0 \mu$m, with a power-law size  index $\alpha = 3.8$ and
     wavelength $\lambda$ = 0.485 $\mu$m.The  curve (marked by 'o'symbols) is the one drawn  using the DDA method, and the  curve  (joined by '+' symbols)  is the one drawn using the Mie theory. }
    \label{pol-485-3.8}
\end{figure}

\begin{figure}
	% To include a figure from a file named example.*
	% Allowable file formats are eps or ps if compiling using latex
	% or pdf, png, jpg if compiling using pdflatex
	\includegraphics[width=\columnwidth]{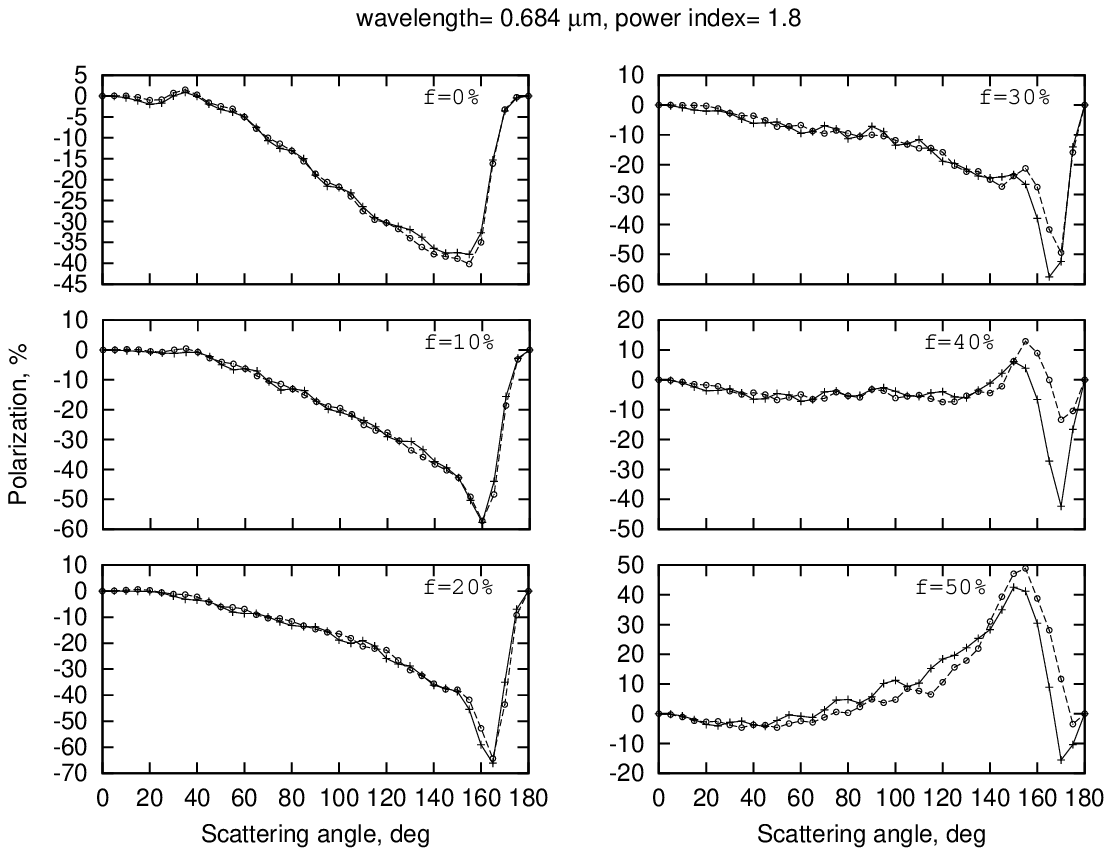}
    \caption{The above figures show the variations of dust polarization with scattering angles expressed in degrees, for various values of dust particle porosities.
    The polarization has been calculated for particles having size range $ 0.01 \mu$m - $1.0 \mu$m, with a power-law size  index $\alpha = 1.8$ and
    wavelength $\lambda$ = 0.684 $\mu$m. The  curve (marked by 'o' symbols) is the one drawn  using the DDA method, and the  curve  (joined by '+' symbols) is the one drawn using the Mie theory.}
    \label{pol-684-1.8}
\end{figure}

\begin{figure}
	% To include a figure from a file named example.*
	% Allowable file formats are eps or ps if compiling using latex
	% or pdf, png, jpg if compiling using pdflatex
	\includegraphics[width=\columnwidth]{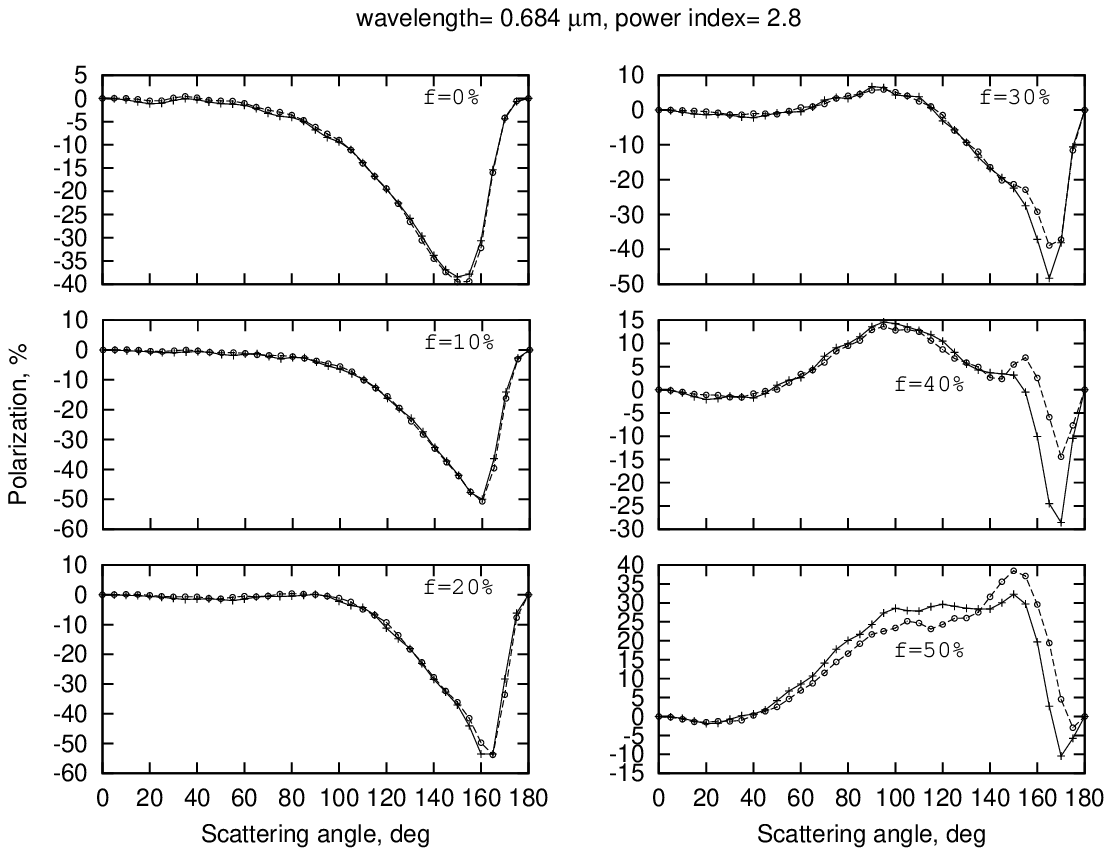}
    \caption{The above figures show the variations of dust polarization with scattering angles expressed in degrees, for various values of dust particle porosities.
    The polarization has been calculated for particles having size range $ 0.01 \mu$m - $1.0 \mu$m, with a power-law size  index $\alpha = 2.8$ and  wavelength $\lambda $ = 0.684 $\mu$m.
    The  curve (marked by 'o' symbols) is the one drawn  using the DDA method, and the  curve  (joined by '+' symbols) is the one drawn using the Mie theory.}
    \label{pol-684-2.8}
\end{figure}

\begin{figure}
	% To include a figure from a file named example.*
	% Allowable file formats are eps or ps if compiling using latex
	% or pdf, png, jpg if compiling using pdflatex
	\includegraphics[width=\columnwidth]{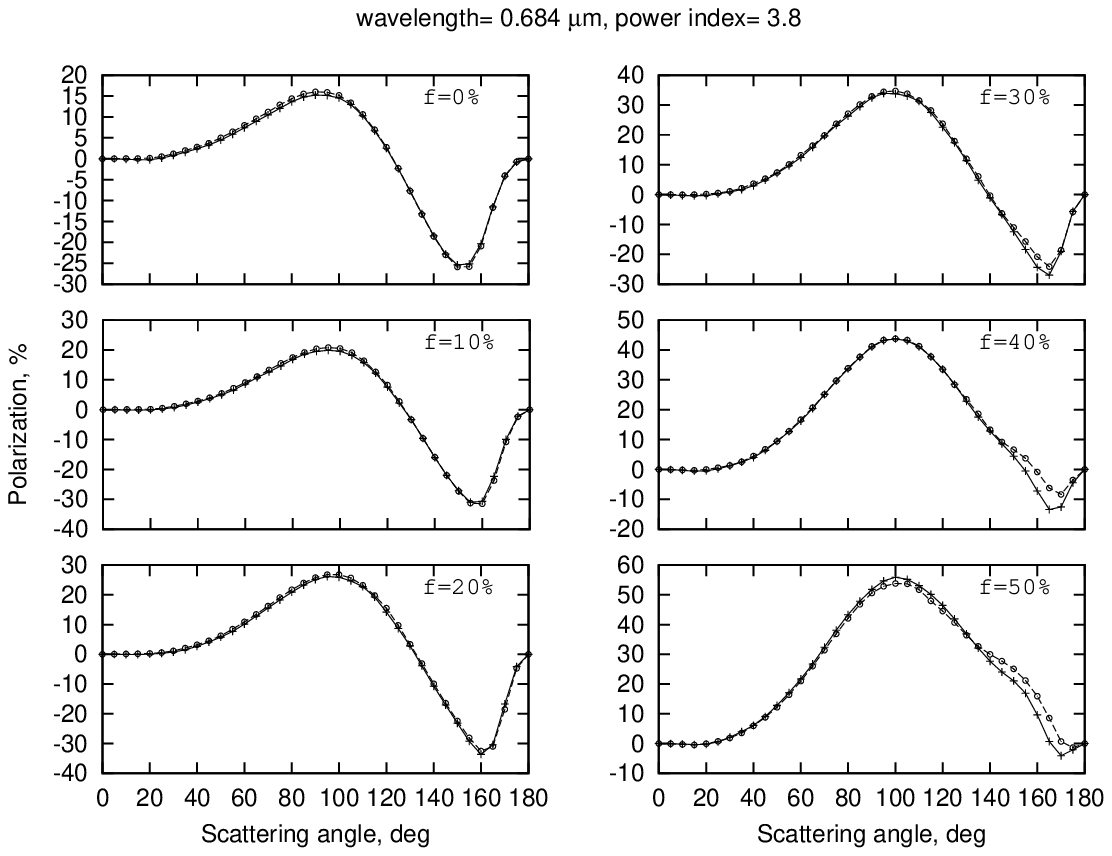}
    \caption{The above figures show the variations of dust polarization with scattering angles expressed in degrees, for various values of dust particle porosities.
    The polarization has been calculated for particles having size range $ 0.01 \mu$m - $1.0 \mu$m, with a power-law size  index $\alpha = 3.8$ and
     wavelength $\lambda$ = 0.684 $\mu$m. The  curve (marked by 'o' symbols) is the one drawn  using  the DDA method, and the  curve  (joined by '+' symbols) is the one drawn
      using  the Mie theory.}
    \label{pol-684-3.8}
\end{figure}

% Example table
\begin{table}
	\centering
	\caption{List of certain porosity values with the corresponding simulated  $P_{max}$ (in $\%$), $P_{min}$ (in $\%$) values and the crossover angle (in degrees) between the positive and the negative polarization branches, from bell-shaped polarization
curves for comet. Only those simulated values are tabulated  which are  close to the observed polarization values. }
	\label{tab:example_table}
	\begin{tabular}{lcccr} % five columns, alignment for each
		\hline
		 $\lambda$ ($\mu$m) & porosity $f$& $P_{max}$ & $P_{min}$ & crossover angle (degrees)\\
		\hline
		0.485 & 40\% (2.8) & 16 & -8 & 160\\
		0.485 & 40\% (3.8) & 45 & -8 & 157\\
        0.684 & 40\% (3.8) & 40 & -8 & 158\\
		\hline

	\end{tabular}
\end{table}

\begin{table}
	\centering
	\caption{The complex Effective Refractive Index $(n' , k')$ values as calculated by using Effective Medium Theory, at two wavelengths for various porosity values. }
	\label{tab:example_table}
	\begin{tabular}{lccr} % four columns, alignment for each
		\hline
		 $\lambda$ ($\mu$m) & porosity $f$\% & $n'$ & $k'$ \\
		\hline
		0.485 & 0 & 1.692 & 0.0492\\
        0.485 & 10& 1.621 &  0.0442\\
        0.485 & 20& 1.549 & 0.0391\\
        0.485 & 30& 1.477 & 0.0338\\
        0.485 & 40& 1.404 & 0.0283\\
        0.485 & 50& 1.331 & 0.0228\\
        0.684 & 0 & 1.672 & 0.0185\\
		0.684 & 10 & 1.603 & 0.0166\\
        0.684 & 20 & 1.533 & 0.0147\\
        0.684 & 30 & 1.463 & 0.0127\\
        0.684 & 40 & 1.392 & 0.0107\\
        0.684 & 50 & 1.322 & 0.0086\\
		\hline

	\end{tabular}
\end{table}

\begin{figure}
	% To include a figure from a file named example.*
	% Allowable file formats are eps or ps if compiling using latex
	% or pdf, png, jpg if compiling using pdflatex
	\includegraphics[width=\columnwidth]{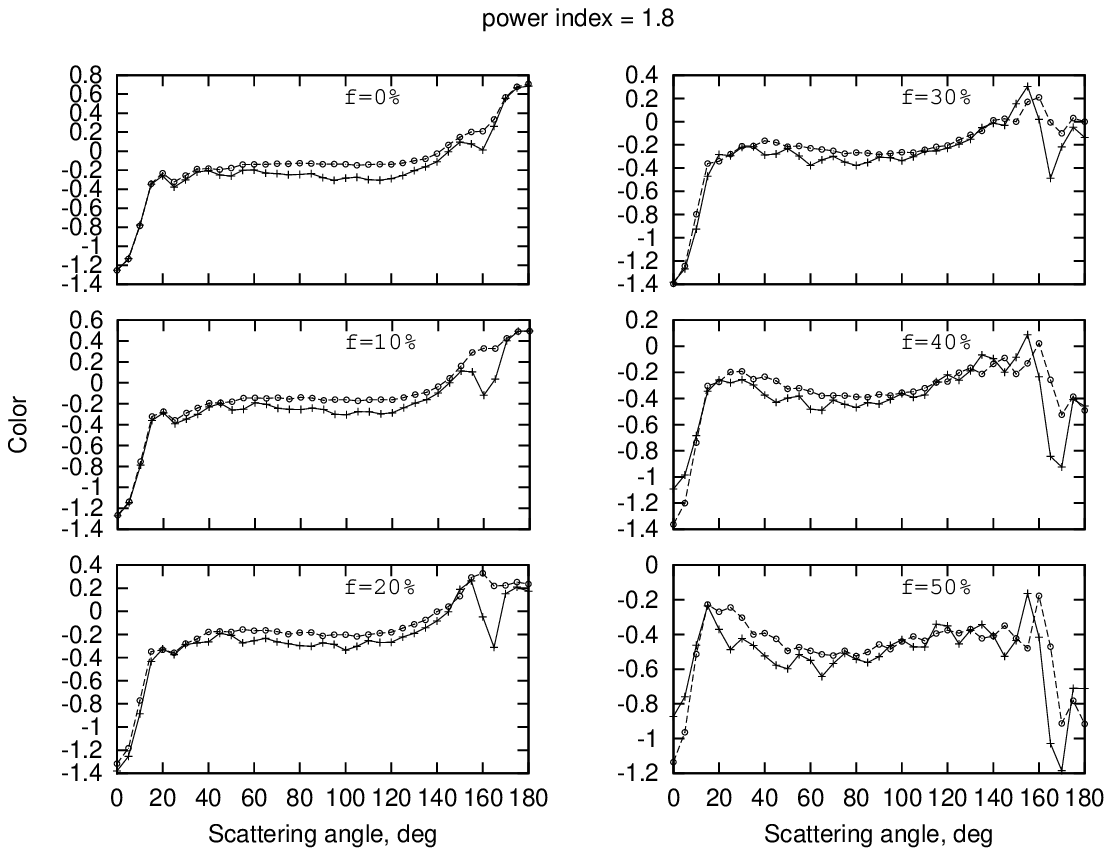}
    \caption{ The above figure shows the variations of dust color with scattering angles expressed in degrees, for various values of dust particle porosities.
    The color has been calculated for particles having size range $0.01 - 1.0 \mu$m , with a power-law size  index $\alpha = -1.8$. The  curve (marked by 'o' symbols) is the one drawn  using the DDA method, and the curve  (joined by '+' symbols) is the one drawn using the Mie theory.
    }
    \label{col-1.8}
\end{figure}

\begin{figure}
	% To include a figure from a file named example.*
	% Allowable file formats are eps or ps if compiling using latex
	% or pdf, png, jpg if compiling using pdflatex
	\includegraphics[width=\columnwidth]{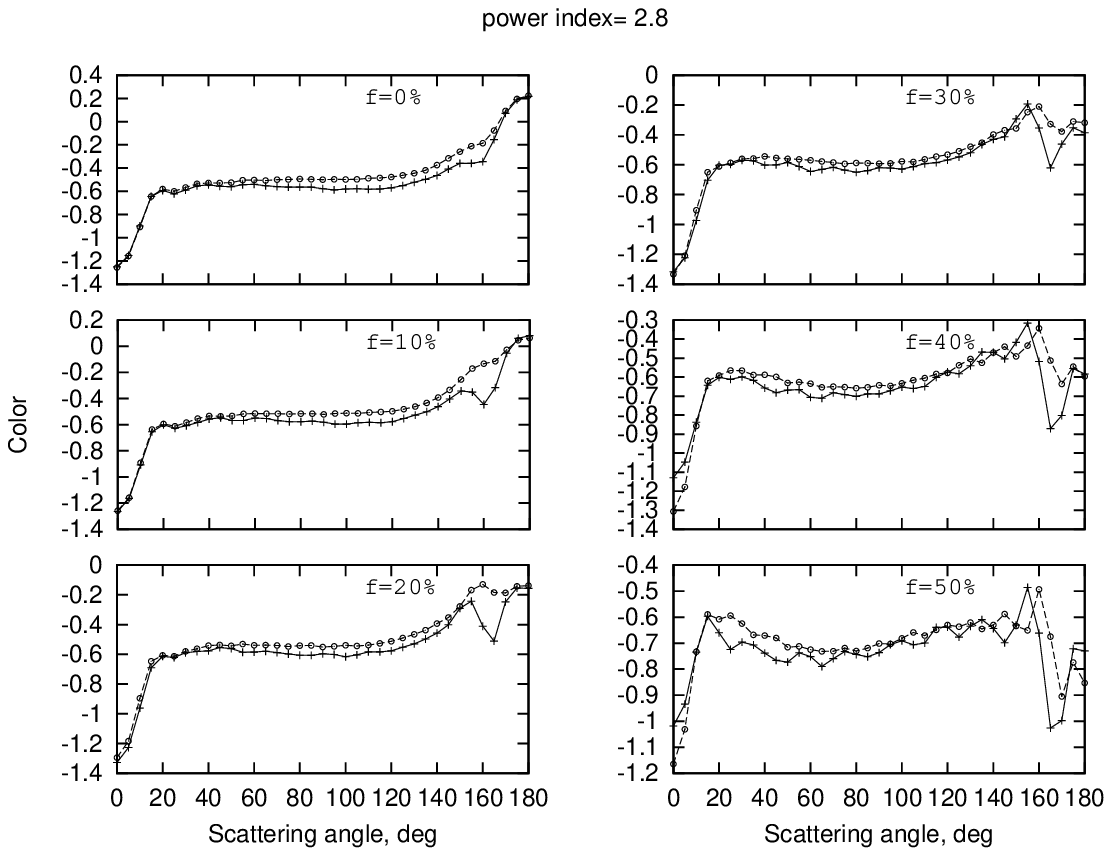}
    \caption{The above figure shows the variations of dust color with scattering angles expressed in degrees, for various values of dust particle porosities.
    The color has been calculated for particles having size range $0.01 - 1.0  \mu$m, with a power-law size  index $\alpha = -2.8$. The  curve (marked by 'o' symbols) is the one drawn  using the  DDA method, and the  curve  (joined by '+' symbols) is the one drawn using the Mie theory.}
    \label{col-2.8}
\end{figure}

\begin{figure}
	% To include a figure from a file named example.*
	% Allowable file formats are eps or ps if compiling using latex
	% or pdf, png, jpg if compiling using pdflatex
	\includegraphics[width=\columnwidth]{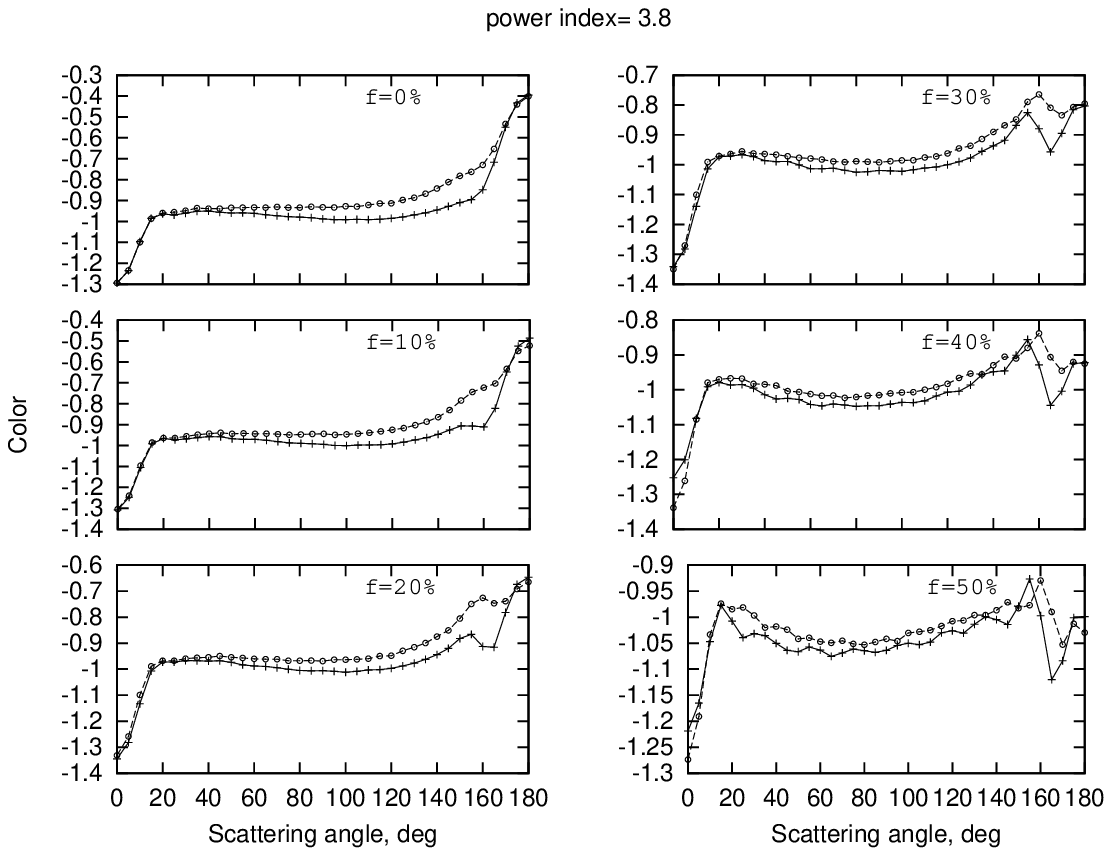}
    \caption{The above figure shows the variations of dust color with scattering angles expressed in degrees, for various values of dust particle porosities.
    The color has been calculated for particles having size range $0.01 - 1.0 \mu$m, with a power-law size  index $\alpha = -3.8$. The  curve (marked by 'o' symbols) is
    the one drawn  using the DDA method, and the  curve  (joined by '+' symbols) is the one drawn using the Mie theory.}
    \label{col-3.8}
\end{figure}

\section{Conclusions}

The aim and main emphasis of the present work, was  to develop a new generic algorithm able to generate realistic
porous aggregates with tunable porosity for a given dust size distribution.
We focused on porous spheres, but generalization to other shapes (e.g. porous ellipsoids) is straightforward.

Using the DDSCAT light scattering code, we showed that such dust particles can generate meaningful polarization values for comets, and hence provide a tool to study in a systematic way the effect of dust porosity on cometary light scattering phenomenon, such as polarization, \emph{color}, etc.

From the analysis of the simulated polarization curves, we find that with the increase in porosity (from $0\%$ to $50\%$) both the maximum and minimum  polarization values change in a non-monotonic way.
Indeed, the negative polarization branch becomes deeper with the increase in porosity and then again it becomes shallower. This phenomenon is true for both the wavelengths $\lambda= 0.485 \mu$m and $= 0.684 \mu$m and with the three values of power law index $\alpha = 1.8, 2.8$ and $3.8$ considered here. \\

More specifically, we have successfully generated polarization values which match grossly with the observed  polarization data for comet Halley.
It appears that the matching with the observed polarization data of comet Halley is better with $f = 40\%$ porosity and $\alpha =  3.8$ radius-distribution power law index.
A sequel work is planned to find the optimum values of the two parameters $f , \alpha$ which realize the least square fitting to the observed data.

On the other hand, the simulated cometary \emph{color} values show only weak dependence on the porosity values  or on the scattering angles (except when close to back-scattering).
At the porosity $f = 40\%$ and power law index $\alpha =3.8$,  the simulated
\emph{color} value $C \simeq -0.6$ does not match with the average observed  \emph{color} value tabulated earlier by Kolokolova et al (1987) which is $ \simeq -0.024$. This discrepancy may come from the underestimation of the upper limit of the particle radius or from the size dependence of single particle $albedo$.

Generally, the simulated \emph{color} values solely appear to be dependent on the power law index of dust size distribution function. Thus the dust \emph{color}  depends on dust particle size and this finding confirms a trend reported by Kolokolova et al (1997) earlier.

At last, we showed that the Effective Medium Theory can only be used qualitatively to understand various features in the polarization or \emph{color} behaviors. If quantitative achievement is required, more detailed theory has to be used, such as the DDA method applied to porous particles as generated by the algorithm given in the present paper.

\section*{Acknowledgements}

R.B. thanks the Assam University, Silchar, for hospitality in the framework of `Global Initiative for Academic Network' (GIAN) programme from MHRD, April 2016. The authors AKS and RV thank Erasmus Mundus - NAMASTE programme for funds to do this collaborative work.

%\section*{References}

\appendix

\section{Some extra material}

If you want to present additional material which would interrupt the flow of the main paper,
it can be placed in an Appendix which appears after the list of references.

%\bsp	% typesetting comment
\label{lastpage}

\begin{thebibliography}{99}
\bibitem{Barabasi 2011}Barab\'asi A.L., Vicsek T., 1990, J. Phys.A, 23, L729

\bibitem{Bertini 2007} Bertini I., Thomas N., Barberi C., 2007, A\&A, 461, 351

\bibitem{Bohren1983}Bohren, C. F., Hofman, D. R., 1983. Absorption and Scattering of Light by Small Particles. Wiley, New York.

\bibitem{Brownlee1985}Brownlee D. E., 1985, Ann. Rev. Earth Planet. Sci., 13, 147

\bibitem{Das 2011}Das H.S., D. Paul, A. Suklabaidya and A. K. Sen,2011, MNRAS 416,94-100

\bibitem{Das et al 2011}Das H.S., Sen A. K., 2011, J. Quant. Spectrosc. Radiative Transfer, 112, 1833

\bibitem{Demers et al 1987}Demers A., Greene D., Hauser C., Irish W., Larson J., Shenker S., Sturgis H., Swinehart D., Terry D. in Proceedings of the 6th Annual ACM Symposium on Principles of Distributes Computing (PODC' 87), 1987. p. 1-12

\bibitem{Draine1994}Draine B.T., Flatau P.J. Discrete-dipole approximation for scattering calculations. J Opt Soc Am A 1994;11:1491-1499.

\bibitem{Draine 2000}Draine B.T. in: Mishchenko M.I., Hovenier, J.W., Travis, L.D., editors. Light Scattering by Nonspherical Particles, Theory, Measurements, and Applications. New York: Academic Press, 2000. p. 131-145

\bibitem{Gupta 2006}Gupta R., Vaidya D. B.,  Bobbie J.S; Chylek Petr, Astrophysics \& Space Science, 301, issue 1, pp. 21-31, 2006

\bibitem{Gustafson and Kolkolova 1999}Gustafson B.S., Kolokolova L., 1999, J. Geophys. Res., 104, 31711

\bibitem{Greenberg 1990}Greenberg J. M.,\& Hage J. I., 1990, ApJ, 361, 260

\bibitem{Hadmcik 2010}Hadamcik,E., Sen,A.K., Levasseur-Regourd,A.C., Gupta,R., Lasue,J., 2010.Astron.Astrophys.517,A86.

\bibitem{Hadmcik 2013}Hadamcik,E., Sen,A.K., Levasseur-Regourd,A.C., Gupta,R., Lasue,J., Botet,R.,2013, Icarus222,774-785.

\bibitem{Hadmcik 2015}Hadamcik,E., Sen,A.K., Levasseur-Regourd,A.C., RoyChoudhury,S., Lasue,J., Gupta, R., Botet,R.,2014. Meteor.Planet.Sci.49(Nr 1 ),36-44.

\bibitem{Hall1949}Hall, J.S. 1949, Sci, 109, 166

\bibitem{Hiltner1949}Hiltner, W.A. 1949, Sci, 109, 165

\bibitem{von Humboldt 1858}von Humboldt, A.1858, Cosmos - a sketch of the physical description of the universe -. reprinted The Johns Hopkins University Press, 1997

\bibitem{Jiang 2006}Jiang, S.Z., editor. Focus on Combustion Research. New York: Nova Science Publishers, 2006

 \bibitem{Jullien 1987} Jullien, R., Botet, R. 1987, Aggregation and Fractal Aggregates, World Scientific Publishing, Singapore

\bibitem{Kimura 2001}Kimura H., 2001, J. Quant. Spectrosc. Radiat. Transfer, 70, 581

\bibitem{Kimura 2006}Kimura H., Kolokolova L., Mann I., 2006, A\&A, 449, 1243

\bibitem{Kolkolova 2007} Kolokolova L., Kimura H., Kiselev N., Rosenbush V., 2007, A\&A, 463,1189

\bibitem{Kolokolova 2001}Kolokolova L , Jockeres K, Gustafson Bo A S, and Gunther L, JOURNAL OF GEOPHYSICAL RESEARCH, 2001, VOL. 106, NO. E5, 10,113-10,127

\bibitem{Kolokolova 1997} Kolokolova L, Jockers K, Chernova G 1997, ICARUS 126, 351-361

\bibitem{Maxwell 1873} Maxwell J.C. 1873 Nature VIII, 437-441

\bibitem{Maxwell 1904} Maxwell-Garnett J.C.1904 Philos Trans R Soc A, 203: 385-420.

\bibitem{Meakin 1983}Meakin P., 1983, J. Colloid Interface Sci., 96, 415

\bibitem{Meaking 1984} Meakin P., 1984, Phys. Rev. A, 29, 997

\bibitem{Mischenko1994}Mishchenko, M. I., \& Travis, L. D. 1994, Opt. Comm., 109, 16

\bibitem{Moreno2007}Moreno, F. , Muñoz, O. , Guirado, D.  and Vilaplana, R., 2007, JQSRT, 106, 348

\bibitem{Mukai 1987} Mukai T, Mukai S, \& Kikuchi S, 1987, Astron. Astrophys., 187, 650

\bibitem{Petrova 2004} Petrova E. V., Tishkovets V. P., Jockers K., 2004, Sol. Syst. Res., 38, 309

\bibitem{Rotundi 2009}Rotundi A., and  other 39 authors., 2015, Science,347,6220, p.3905-1

\bibitem{Rayleigh 1871}Lord Rayleigh. 1871, Philosophical Magazine,41, p.107-120

\bibitem{Seizinger 2013}Seizinger A., Kley W., 2013,  Astron \& Astrophys, A65

\bibitem{Sen 1991}Sen A .K, Deshpande M R, Joshi U C, Rao N K, Raveendran A V, 1991,  Astron \& Astrophys, 242, 496

\bibitem{Shen 2008}Shen Y., Draine B. T., Johnson E. T., 2008, ApJ, 689, 260

\bibitem{Shen 2009}Shen Y., Draine B. T., Johnson E. T., 2009, ApJ, 696, 2126

\bibitem{Tishkovets 2004}Tishkovets V. P., Petrova E.V., Jockers K., 2004, J. Quant. Spectrosc. Radiat. Transfer, 86, 241

\bibitem{Taflove 2005}Taflove A., Hagness S.C., Computational Electrodynamics: the Finite-Difference Time-Domain Method, 3rd ed. Norwood: Artech House, 2005

\bibitem{Hulst1957} van de Hulst H.C., Light scattering by small particles, Wiley, New York, 1957

\bibitem{Vilaplana2006}Vilaplanaa R.,  Moreno F., Molina A., 2006, JQSRT,100,415

\bibitem{Vilaplana2011}Vilaplana R., Luna R., Guirado, D., 2011, JQSRT,112, 1838

\bibitem{Vaidya 2007} Vaidya D.B, Gupta R, Snow T.P, 2007 MNRAS 379, 791-800

\bibitem{Voshchinnikov 2007} Voshchinnikov N.V., Videen G., and Henning T., 2007 Applied Optic, vol 46, Issue 19, 4065

\bibitem{Wang 1995}Wang C. Y., Liu P. L., Bassingthwaighte J. B., 1995, J. Phys.A Math. Gen., 28, 2141

\bibitem{Waterman 1971}Waterman, P. C. 1971, Phys. Rev. D, 3, 825

\bibitem{Wurm 1998}Wurm G., Blum J., 1998, Icarus, 132, 125

\bibitem{Xing 1998}Xing Z., Hanner M. S., 1997, A\&A, 324, 805

\end{thebibliography}
\end{document}